\documentclass[nofootinbib,superscriptaddress]{revtex4}
\usepackage{graphicx}
\usepackage{subfigure}
\usepackage{dcolumn}
\usepackage{bm}

\begin{document}
%


\title{On vacuum gravitational collapse in nine dimensions}

\author{P. Bizo\'n}
\affiliation{\small{M. Smoluchowski Insitute of Physics,
Jagiellonian University, Krak\'ow,
 Poland}}
\author{T. Chmaj}
\affiliation{\small{H. Niewodnicza\'nski Institute of Nuclear
Physics, Polish Academy of Sciences, Krak\'ow, Poland}}
\affiliation{\small{Cracow University of Technology, Krak\'ow,
    Poland}}
\author{A. Rostworowski}
\affiliation{\small{M. Smoluchowski Insitute of Physics,
Jagiellonian University, Krak\'ow,
 Poland}}
\author{B. G. Schmidt}
 \affiliation{\small{Max-Planck-Institut
f\"ur Gravitationsphysik, Albert-Einstein-Institut, Golm, Germany}}
\author{Z. Tabor}
\affiliation{\small{Department of Biophysics, Jagellonian
University, Krak\'ow, Poland}}

\date{\today}
\begin{abstract}
We consider the vacuum gravitational collapse for cohomogeneity-two
solutions of the nine dimensional Einstein equations. Using combined
numerical and analytical methods we give evidence that within this
model the Schwarzschild-Tangherlini black hole is asymptotically
stable. In addition, we briefly discuss the critical behavior at the
threshold of black hole formation.
\end{abstract}

\maketitle
\noindent \emph{\textbf{Introduction and setup:}}
 Over the past few years there has been a surge of interest in
higher dimensional gravity, motivated by several reasons. The main
reason comes from the new brane-world scenario in string theory
according to which we live on a three dimensional surface (called a
brane) in a higher dimensional spacetime \cite{rs}. In order to
understand the phenomenology of this scenario and its experimental
predictions, like production of black holes in the next generation
of colliders, it becomes important to study solutions of
 Einstein's equations in more than four dimensions. Another reason, closer to our own motivation, has
  nothing to do with string
 theory and comes from general relativity itself. Viewing the
 dimension of a spacetime as a parameter of the theory is helpful in
 understanding which features of general relativity depend crucially on our world being four
 dimensional and which ones hold in general.
 Last, but not
 least, extra dimensions are fun.

In a recent paper \cite{bcs} some of us showed that in five
spacetime dimensions one can perform a consistent cohomogeneity-two
symmetry reduction of the vacuum Einstein equations which -- in
contrast to the spherically symmetric reduction -- admits time
dependent asymptotically flat solutions.  The key idea was to modify
the standard spherically symmetric ansatz by replacing the round
metric on the three-sphere with the homogeneously squashed metric,
thereby breaking the $SO(4)$ isometry to $SO(3)\times U(1)$. In this
way the squashing parameter becomes a dynamical degree of freedom
and the Birkhoff theorem is evaded. This model provides a simple
theoretical setting for studying the dynamics of gravitational
collapse in vacuum, both numerically \cite{bcs} and analytically
\cite{dh}.

As mentioned in \cite{bcs}, similar models can be formulated in
higher $D=n+2$ dimensions as long as the corresponding sphere $S^n$
admits a non-round homogeneous metric, i. e. there exists a proper
subgroup of the orthogonal group $SO(n+1)$ which acts transitively
on $S^n$. According to the classification given by Besse
\cite{besse}, such transitive actions exist on all odd dimensional
spheres. For example, the group $SU(n+1)$ acts transitively on
$S^{2n+1}$ and the group $Sp(n+1)$ acts transitively on $S^{4n+3}$.
It is natural to ask whether the properties of gravitational
collapse found in \cite{bcs} are typical for this class of models or
whether new phenomena appear in higher dimensions.  A systematic
analysis of this question appears hopeless in view of the fact that
the number of degrees of freedom (squashing parameters) grows
quickly with dimension, thus it seems useful to look at specific
examples  to get better understanding. In this note we consider a
model with one degree of freedom which describes the squashing of
the seven sphere. More concretely, we regard $S^7$ as the coset
manifold $Sp(2)/Sp(1)\simeq SO(5)/SO(3)$, or equivalently, as the
$S^3$ bundle over the $S^4$ base space with $SO(5)\times SO(3)$
invariant metric. In the past, such the squashed seven sphere has
attracted a great deal of attention in the context of eleven
dimensional supergravity \cite{dnp}.

We parametrize the metric in the following way
\begin{equation}\label{metric}
    ds^2= - A e^{-2\delta} dt^2 + A^{-1} dr^2 + r^2
    d\Omega_7^2,
\end{equation}
where $d\Omega_7^2$ is the metric on the unit squashed $S^7$
\cite{adp}
\begin{equation}
d\Omega_7^2 = \frac{1}{4} e^{3B} (d\mu^2+e_i^2)+\frac{1}{4} e^{-4B}
E_i^2,
\end{equation}
 $B$, $A$, and $\delta$ are  functions of $t$ and $r$,
\begin{equation}
    e_i=\frac{1}{2} \sin{\mu}\, (\sigma_i-\tilde\sigma_i),\quad
    E_i=\cos^2{\frac{\mu}{2}}\, \sigma_i+\sin^2{\frac{\mu}{2}}\, \tilde\sigma_i,
\end{equation}
and $\sigma_i$ and $\tilde\sigma_i$ are two sets of left-invariant
one-forms on $SU(2)$
\begin{equation}
  \sigma_1 = \cos{\psi} d\theta+\sin{\psi}\sin{\theta} d\phi,
  \quad
  \sigma_2 = -\sin{\psi} d\theta+\cos{\psi}\sin{\theta} d\phi,
  \quad
  \sigma_3 = d\psi+\cos{\theta} d\phi,
\end{equation}
with $\tilde \sigma_i$  given by identical expressions in terms of
$(\tilde\psi,\tilde\theta,\tilde\phi)$.
 In this ansatz the $SO(8)$ isometry of the round $S^7$ is broken
 to $SO(5)\times SU(2)$. This is a special case of a more general ansatz for which the $S^3$
 fibers are allowed to be squashed themselves \cite{cgp}.

Substituting the ansatz (\ref{metric}) into the vacuum Einstein
equations, $R_{\alpha\beta}=0$, we get  equations of motion for the
functions $A(t,r), \delta(t,r)$ and $B(t,r)$ (in the following we
use overdots and primes to denote $\partial_t$ and $\partial_r$,
respectively)
\begin{eqnarray}\label{mom}
  A' \!\!&=&\!\! - \frac{6 A}{r} +\frac{1}{7r} \left(48 e^{-3B}-12 e^{-10 B}+6 e^{4 B}\right)-
  3 r
 \left( e^{2\delta} A^{-1} {\dot B}^2 + A {B'}^2\right)\\
    \dot A \!\!&=& \!\!-  6 r A \dot B B',\\
    \delta' \!\!&= &\!\!- 3 r \left(e^{2\delta} A^{-2}{\dot B}^2 +
    B'^2\right),
\end{eqnarray}
\vspace{-0.5cm}
\begin{equation}\label{wave}
\left(e^{\delta} A^{-1} r^7 {\dot B}\right)^{\cdot} -
\left(e^{-\delta} A r^7 B'\right)' + \frac{4}{7} e^{-\delta} r^5
\left(6 e^{-3B}-e^{4B}-5 e^{-10B} \right)=0.
\end{equation}
It follows immediately from the above equations that if $B=0$ (no
squashing), then $\delta=0$ and either $A=1$ (Minkowski), or
$A=1-r_h^6/r^6$ (Schwarzschild-Tangherlini), where $r_h$ is the
radius of the black hole horizon. We showed in \cite{bcs} that the
analogous two solutions in five dimensions play the role of generic
attractors in the evolution of regular asymptotically flat initial
data (small and large ones, respectively) and the transition between
these two outcomes of evolution exhibits the type II discretely
self-similar critical behavior. The aim of this note is to show that
these properties  are also present in nine dimensions which suggests
that they are general features  of vacuum gravitational collapse for
this class of models.
\vskip 0.2cm \noindent \emph{\textbf{Linear stability and
quasinormal modes:}}
Before presenting  numerical evidence for the nonlinear stability of
the Schwarzschild-Tangherlini solution, we want to discuss the
results of linear perturbation theory. Linearizing equations
(\ref{mom})-(\ref{wave}) around the Schwarzschild-Tangherlini
solution we obtain the linear wave equation for the perturbation
$\delta B(t,r)$
\begin{equation}
\label{pert} \ddot{\delta\!B} - \frac{1}{r^7} A_0 (r^7 A_0
{\delta\!B'})' + \frac{16 A_0}{r^2} \delta\!B=0, \qquad
A_0=1-\frac{1}{r^6},
\end{equation}
where we have used the scaling freedom to  set the radius of the
horizon $r_h=1$. Introducing the tortoise coordinate $x$ defined by
$dx/dr\,=\,A_0^{-1}$,
and substituting $\delta B(t,r)=e^{-i k t} r^{-7/2} u(x)$ into
(\ref{pert}) we get the Schr\"odinger equation on the real line
$-\infty < x < \infty$
\begin{equation}\label{schr}
    -\frac{d^2 u}{dx^2} + V(r(x)) u = k^2 u,\quad
    V(r)=\frac{1}{4}
    \left(1-\frac{1}{r^6}\right)\left(\frac{99}{r^2}+\frac{49}{r^8}\right).
\end{equation}
This equation is a special case (corresponding to the gravitational
tensor perturbation with $l=2$) of the master equation for general
perturbations of the higher dimensional Schwarzschild-Tangherlini
solutions derived independently by Gibbons and Hartnoll \cite{gh}
and Ishibashi and Kodama \cite{ik}. Since the potential in
(\ref{schr}) is everywhere positive, there are no bound states,
which implies that the Schwarzschild-Tangherlini black hole is
linearly stable.

We computed quasinormal modes, i.e. solutions of equation
(\ref{schr}) satisfying the outgoing wave boundary conditions $u\sim
e^{\pm i k x}$ for $x\rightarrow \pm\infty$,
using Leaver's method of continued fractions \cite{Leaver85,
Leaver90}. Substituting
\begin{equation}
\label{u(r)} u(x(r)) = \left( \frac {r-1} {r+1} \right)^{-i k/6}
e^{i k r} \sum_{n=0}^{\infty} a_n \left( \frac {r-1} {r} \right)^n,
\qquad a_0 = 1,
\end{equation}
into (\ref{schr}) we get a 12-term recurrence relation for the
coefficients $a_n$.
 The prefactor in (\ref{u(r)}) fulfills the outgoing wave
boundary conditions  both at the horizon and at infinity, hence the
quasinormal frequencies are given by the discrete values of
 $k$ for which the series in (\ref{u(r)}) converges at $r=\infty$.
Using Gaussian elimination \cite{Leaver90} we reduced the 12-term
recurrence relation to a 3-term one. According to Leaver
\cite{Leaver85}, the convergence of the series $\sum_{n=0}^{\infty}
a_n$ is equivalent to a condition under which the 3-term recurrence
relation has a minimal solution. Such the condition, formulated in
terms of continued fractions, yields the transcendental eigenvalue
equation which has to be solved numerically.  The full details of
this computation will be described elsewhere.  Here, in order to
interpret numerical results we need only to know that the
fundamental quasinormal mode has the frequency $k_0=3.4488-0.8601 i$
(in units $r_h^{-1}$). This mode is expected to dominate the process
of ringdown. The higher modes are damped much faster (for example,
the first overtone has the frequency $k_1=2.7548-2.6116 i$) so in
practice they play no role in the dynamics.
\vskip 0.2cm
 \noindent \emph{\textbf{Numerical results:}}
Using the same finite-difference code as in \cite{bcs} we solved
equations (5)-(8) numerically for several families of regular
initial data. We
 found that the overall picture of dynamics of gravitational
collapse looks qualitatively the same as in five dimensions, that
is, we have dispersion to the Minkowski spacetime for small data and
collapse to the Schwarzschild-Tangherlini black hole for large data
(see Fig. 1a). In the latter case we looked in more detail at the
asymptotics of this process. We found that
 at some intermediate times the solution
settling down to the Schwarzschild-Tangherlini black hole is well
approximated outside the horizon by the least damped quasinormal
mode. This is shown in Fig. 1b.

\begin{figure}[h]
\centering \subfigure[\normalsize{ Formation of a black hole}]{
\label{fig:subfig:a}
\includegraphics[width=0.47\textwidth]{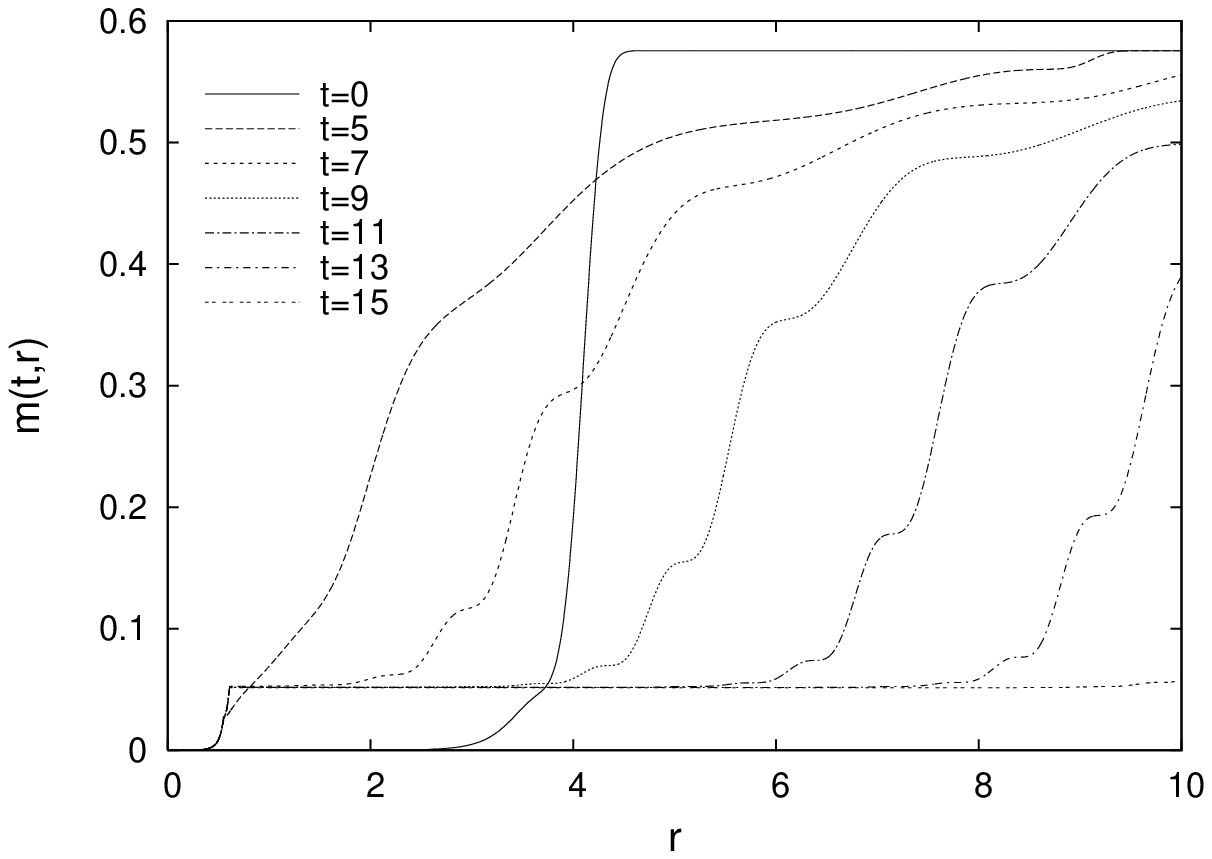}}
\subfigure[\normalsize{ Quasinormal ringing}]{ \label{fig:subfig:b}
\includegraphics[width=0.47\textwidth]{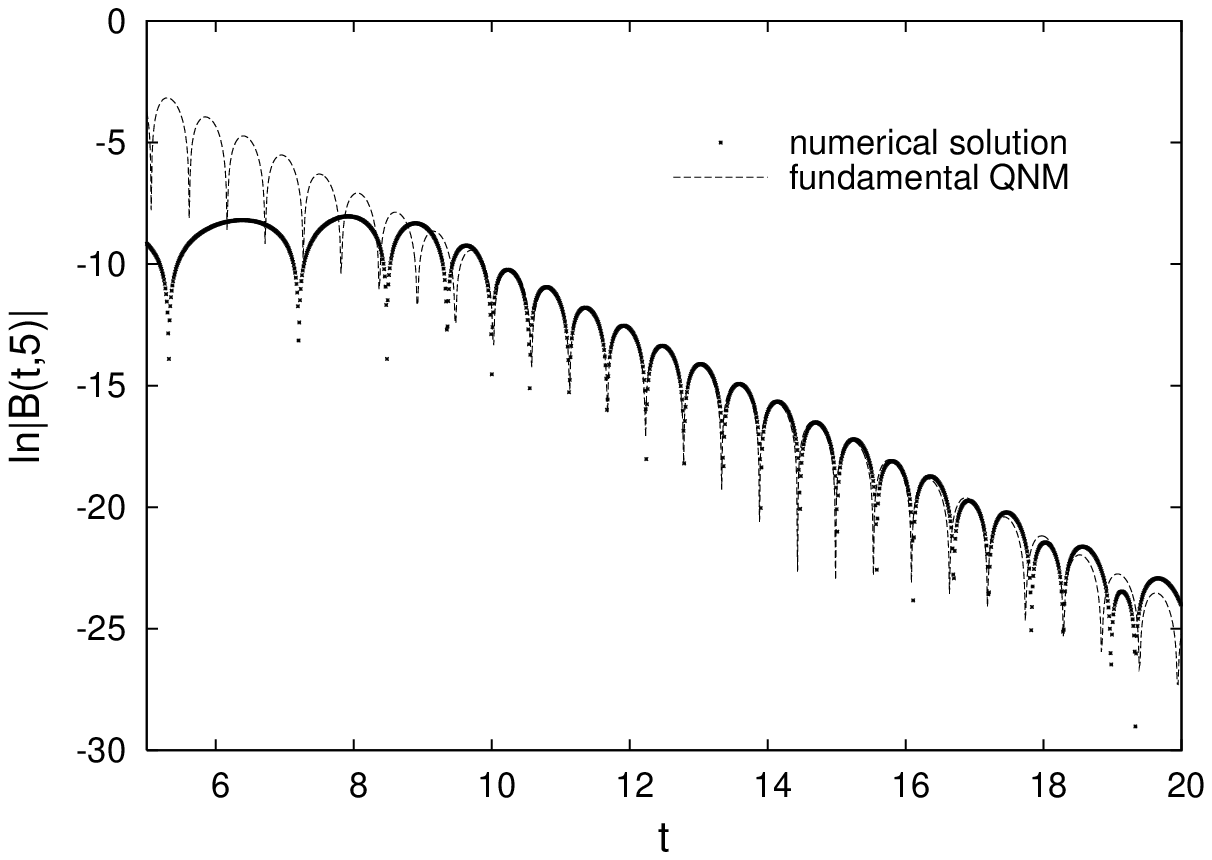}}
\caption{\small{(a) For a black-hole  solution we plot a  series of
late-time snapshots of the mass function $m(t,r)$. The initial mass
function (solid line) has the form of a kink asymptoting the total
mass $m(\infty)=0.575$. During the evolution the horizon develops at
$r_h=0.606$ which corresponds to the black hole mass
$m_{BH}=r_h^6=0.05$. Thus, only a small fraction of the total mass
gets trapped inside the horizon while the remaining mass is being
radiated away to spatial infinity, as is clearly seen from the plot.
(b)
 We plot the time series $\ln|B(t,r_0)|$ at $r_0=5$ for the same solution as in (a),
  and superimpose the fundamental
  quasinormal mode with frequency
$k_0=(3.4488-0.8601\, i)/r_h$. On the time interval $10<t<16$ we get
very good agreement between these two curves which confirms a
well-known fact that quasinormal modes encode an intermediate time
behavior of solutions at a fixed point in space \cite{ks}.}}
\label{fig:subfig}
\end{figure}
\samepage
\begin{figure}[h]
\centering \subfigure[\normalsize{ Discrete self-similarity}]{
\label{fig:subfig:a}
\includegraphics[width=0.47\textwidth]{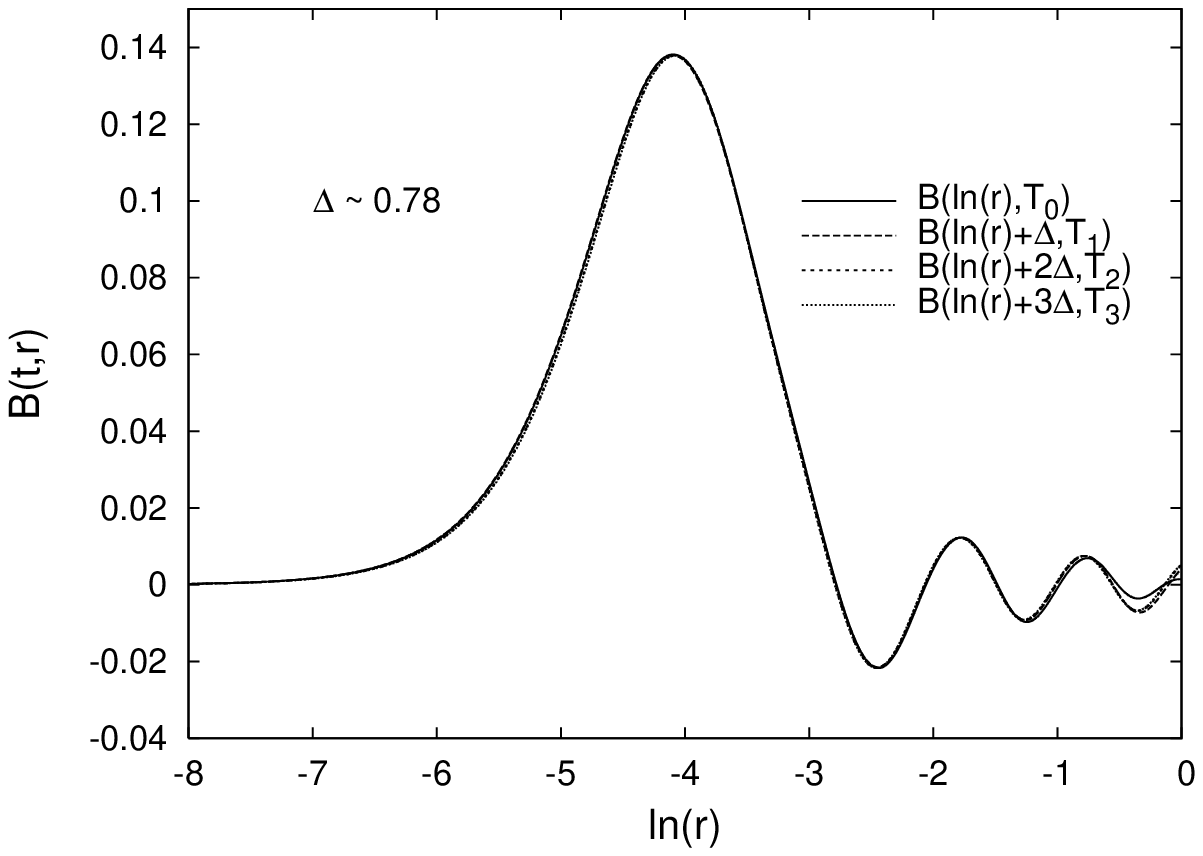}}
\subfigure[\normalsize{ Black-hole mass scaling }]{
\label{fig:subfig:b}
\includegraphics[width=0.47\textwidth]{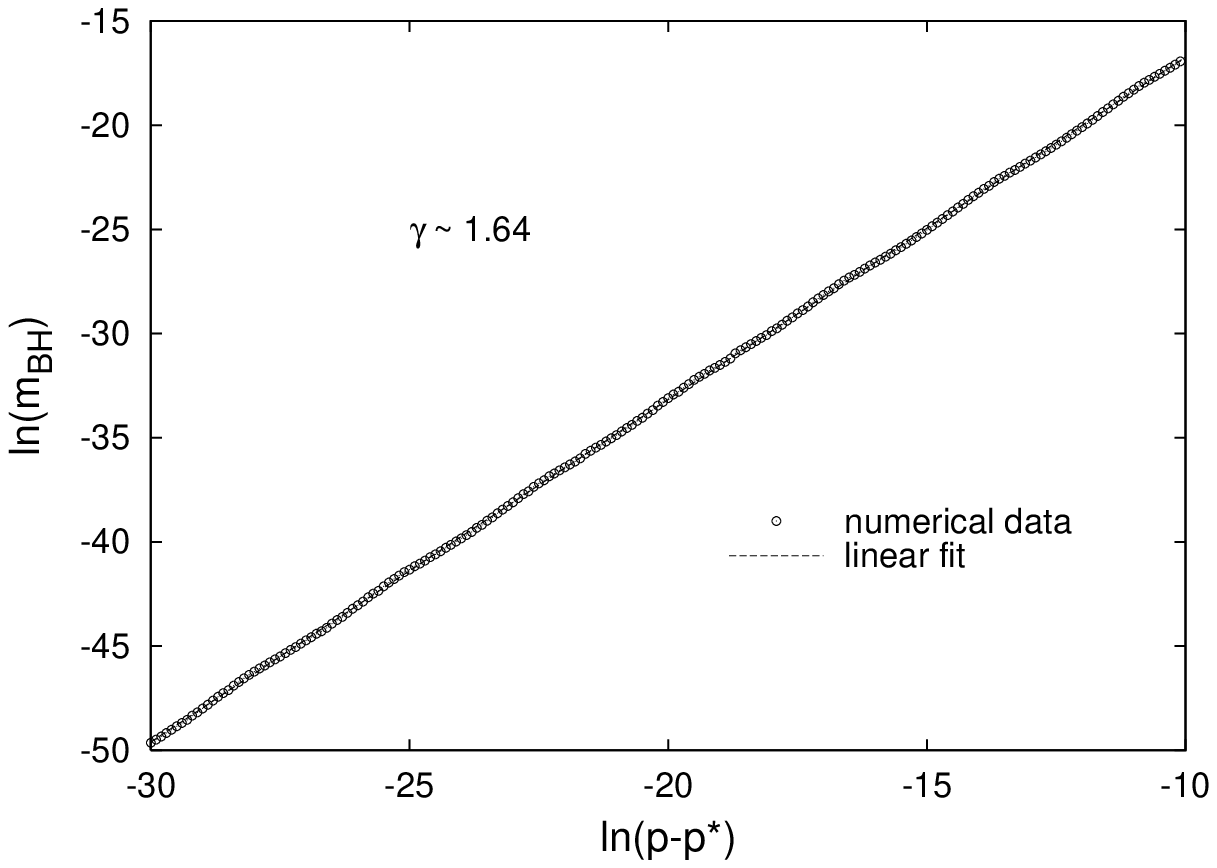}}
\caption{\small{(a) For a near-critical evolution we plot the
function $B$ at some late central proper time $T_0$ and superimpose
the next three echoes (at times $T_n$) shifted by $\ln(r)\rightarrow
\ln(r)+n\Delta$. Minimization of the discrepancy between the
profiles yields $\Delta\approx 0.78$. (b) For supercritical
solutions the logarithm of black hole mass $m_{BH}$ is plotted
versus the logarithmic distance to criticality. A fit of the power
law $\ln(m_{BH})=\gamma \ln(p-p^*) + const$ yields $\gamma\approx
1.64$. The small wiggles around the linear fit are the imprints of
discrete self-similarity; their  period is equal to $2.87$  in
agreement with the theoretical prediction $6\Delta/\gamma$. } }
 \label{fig:subfig}
\end{figure}
\newpage
 \noindent
 At the
threshold for black hole formation we observe the type II discretely
self-similar critical behavior with the echoing period
$\Delta\approx 0.78$ and the black-hole mass scaling law with the
universal exponent $\gamma\approx1.64$. This is shown in Fig.~2.
Note that in eight space dimensions, mass has the dimension of
$length^6$, hence $\gamma=6/\lambda$, where $\lambda$ is the
eigenvalue of the growing mode of the critical solution. Strangely
enough, the product $\Delta \lambda\approx 2.85$ is approximately
the same (up to numerical errors) as in five dimensions \cite{bcs}.
We do not know whether there is any deeper meaning behind this
numerical coincidence.
\vskip 0.2cm \noindent \emph{\textbf{Final remarks:}}
 In this note we have focused on similarities between the models in five and nine
dimensions, however at the end we would like to mention two
interesting qualitative differences which in our opinion are worth
investigation.

The first difference is the existence of the second static solution
 corresponding to the  non-round homogeneous Einstein metric on $S^7$
\begin{equation}\label{SqS}
 B=\frac{\ln{5}}{7},\quad
\quad A=9\cdot 5^{-10/7} \left(1-\frac{r_h^6}{r^6}\right) , \quad
\delta=0.
\end{equation}
 This solution is asymptotically conical so it
does not participate in the dynamics of asymptotically flat initial
data. It would be interesting to study the evolution of
non-asymptotically flat initial data and determine a dynamical role
of the solution (\ref{SqS}), as well as other known explicit
solutions, like for example the Spin(7) solution \cite{gpp}.

 The second difference is the lack of monotonicity of the mass
function $m(t,r)$, defined by $A = 1-m(t,r)/r^6$.
 From the hamiltonian constraint (5) we obtain
 \begin{equation}\label{mass}
 m' = 3 r^7 \left(e^{2\delta} A^{-1} {\dot B}^2 + A {B'}^2\right)+\frac{1}{7}
 r^5
 \left(42-48 e^{-3B}+12
e^{-10B}-6 e^{4B}\right).
\end{equation}
The potential term on the right hand side of this equation has the
local minimum equal to zero at $B=0$, the local maximum at
$B=\ln{5}/7$, and is negative for large positive values of $B$.
Thus, the mass density  may be locally negative and, indeed, it is
easy to construct initial data with large regions of negative mass
density. This suggests a possibility of violating the weak cosmic
censorship (note that monotonicity of mass is essential in the
Dafermos and Holzegel proof \cite{dh}). Although our preliminary
numerical attempts failed to produce a counterexample, i.e. a
generic naked singularity, this problem deserves more systematic
investigation. We should stress that despite locally negative mass
density the total mass $m(\infty)$ is guaranteed to be nonnegative
by  the positive mass theorem in higher dimensions \cite{lohk}. It
would be interesting to prove this fact in our model in an
elementary manner. In particular, for time-symmetric initial data it
follows from equation (\ref{mass}) and the requirement of regularity
at the origin, $m(0)=0$, that
\begin{equation}\label{massint}
  m(\infty)= e^{-\int_0^{\infty} 3 r B'^2 dr} \int_0^{\infty} e^{\int_0^{r} 3 \rho B'^2
  d\rho} \left(3 r^7 {B'}^2 +\frac{1}{7}
 r^5
 (42-48 e^{-3B}+12
e^{-10B}-6 e^{4B})\right) dr.
\end{equation}
 We challenge the readers to show that this integral is nonnegative
for all "reasonable" functions $B(r)$ which vanish at the origin and
have compact support (or fall off faster than $r^{-3}$ at infinity).
Finally, we remark that there is an analogous problem for the
black-hole boundary condition, $m(r_h)=r_h^6$, where one wants to
prove the Penrose inequality $m(\infty)\geq m(r_h)$.
\vskip 0.2cm \noindent \emph{\textbf{Acknowledgments:}} PB is
grateful to Gary Gibbons for discussions, remarks and sharing his
notes. PB acknowledges the hospitality of the Isaac Newton Institute
in Cambridge where this work was initiated. This research was
supported in part by the Polish Ministry of Science grant no.
1PO3B01229.

\end{document}